\documentclass[aps,prd,showpacs,floatfix,preprintnumbers,nofootinbib,amssymb,amsmath]{revtex4}
\usepackage{graphicx,bm,color}

\begin{document}

\title{{Reply to Comment on the paper ``Energy Loss of Charm 
Quarks in the Quark-Gluon Plasma: Collisional vs Radiative''
by Mishra et al}}

\author{Munshi G. \surname{Mustafa}}
\thanks{On leave of absence from Theory Group, Saha Institute of 
Nuclear Physics, 
1/AF Bidhan Nagar, Kolkata 700 064, India}
\email{munshigolam.mustafa@saha.ac.in}

\affiliation{ Department of Physics, McGill University, 3600 
University Street, Montreal, Canada H3A 2T8}

\author{Markus H. \surname{Thoma}}
\email{mthoma@mpe.mpg.de}

\affiliation{Max-Planck-Institut f\"ur extraterrestrische Physik,
P.O. Box 1312, 85741 Garching, Germany}

\vspace{0.2in}

\begin{abstract}

The comments raised in Ref.~\cite{mmp} by Mishra et al aim at two papers
contained in Ref.~\cite{mt}. We show that those comments on Ref.~\cite{mt} 
pointed out by Mishra et al in Ref.~\cite{mmp} are not relevant and the 
concept used in Ref.~\cite{mt} is consistent and in compliance with the 
classical approximation of the transport coefficients~\cite{mgm}. 
We would also like to note that most of the 
comments in Ref.~\cite{mmp} were meant for light quarks,  
but are not even appropriate for heavy quarks. 

\end{abstract}
\maketitle

\vspace{0.2in}

The comments of Mishra et al in Ref.~\cite{mmp} are organised as a list
of observations (``lacunas''). Here we reply to these observations one by one.

\vspace{0.2in}

{\underline{\it Lacuna 1: Notations}}

\vspace{0.2in}

The notations and/or expressions can easily be understood depending upon
the  physical considerations (massless or massive quarks) they are applied to.
In the case of high momenta (at least 
$p> 5$ GeV/c), there is, in fact, no significant difference between $E$ 
and $p$, for a charm quark jet.

\vspace{0.2in}

{\underline{\it Lacuna 2: Fokker-Planck (FP) Equation}}

\vspace{0.2in}
 
We begin with the one-dimensional FP equation~\cite{mt,mgm} as
\begin{equation}
\frac{\partial}{\partial t}D(p,t) = \frac{\partial}{\partial p}
[\Gamma_1(p) D(p,t)] \ + \  \frac{\partial^2}{\partial^2 p}[{\Gamma_2}(p) 
D(p,t)] , \label{fp0}
\end{equation}
that describes the evolution of the momentum distribution, $D(p,t)$, 
in a domain $(-\infty \leq p \leq \infty)$,
at a given time $t \ (0\leq t\leq \infty)$, 
of a test particle undergoing Brownian motion. 
$\Gamma_1(p)$ and $\Gamma_2(p)$ are known as the moments or the FP 
coefficients or the transport coefficients. Usually $\Gamma_1(p)$ is 
related to the collisions whereas $\Gamma_2(p)$ to the momentum diffusion 
in the medium when a test particle undergoes Brownian motion.

We now outline the classical approximations~\cite{mt,mgm} for the
transport coefficients in which the drag force, ${\cal A}$, is assumed to 
be related to the collisional energy loss, $-dE/dL$, as
\begin{equation}
\Gamma_1(p)=-\frac{dE}{dL} \approx p {\cal A}(p) \approx p{\cal A} ,  
 \label{drag}
\end{equation}
where ${\cal A} = \langle {\cal A}(p) \rangle = \langle - 
\frac{1}{p}\frac{dE}{dL}\rangle$,  
and the diffusion coefficient is related to the drag as
\begin{equation}
\Gamma_2(p) \approx p {\cal A}(p) p \approx T {\cal A }(p) p \approx 
T \langle -\frac{dE}{dL} \rangle \equiv {\cal D}_F.
\label{diffusion}
\end{equation}

Within this approximation the drag, ${\cal A}$, and the diffusion, 
${\cal D}_F$, coefficients are momentum independent. 
Now one can write 
the FP equation in (\ref{fp0}) as \cite{mgm}
\begin{equation}
\frac{\partial}{\partial t}D(p,t) = {\cal A}\frac{\partial}{\partial p}
[pD(p,t)] \ + \ {\cal D}_F \frac{\partial^2}{\partial^2 p}D(p,t) .
\label{fp}
\end{equation}
The authors of Ref.~\cite{mmp} 
also agree to the above form (see, e.g, eq. (C3) of Ref.~\cite{mmp}), which
implies that they, obviously, accept the momentum independence 
approximation of ${\cal A}$ and ${\cal D}_F$. 

We now quote the solution of the FP equation (\ref{fp}) for a given 
time $t$ as obtained in Ref.~\cite{mt} 
%using the classical drag and 
%diffusion approximations as stated above, that reads
%\begin{eqnarray}
%D(p,t) \, 
%&=& 
%\, \frac{1} 
%{\sqrt {\pi \, \left (4 \, \int^t {\cal D}_F(t') \, \exp \left [ 
%\, 2\int^{t'}{\cal A}(t'')\, dt'' \right ] \, dt' \right ) \, 
%\left [\exp \left (-2 \int^{t'}{\cal A}(t'')\, dt'' \right ) \right ]}} 
%\nonumber \\
%&\times& \, \exp \left[-\frac{\left (p-p_0\, e^{-\int^t{\cal A}(t')\, dt'}
%\right )^2} 
%{\left (4 \, \int^t {\cal D}_F(t') \, \exp \left [ 
%\, 2\int^{t'}{\cal A}(t'')\, dt'' \right ] \, dt' \right ) \, 
%\left [\exp \left (-2 \int^{t'}{\cal A}(t'')\, dt'' \right ) \right ]} 
%\right ]  \, \, . \label{solmom}
%\end{eqnarray}
%For relativistic particles, $p=E$, (\ref{solmom}) can be written as
\begin{eqnarray}
D(p,t) &=& \frac{1}{\sqrt{\pi\, {\cal W}(t)}} \, \exp \left [ 
- \frac{\left (p-p_0\, e^{-\int^t_0{\cal A}(t') \, dt'} \right )^2}
{{\cal W}(t)} \right ] \, \, , \label{solmom} 
\end{eqnarray}
where ${\cal W}(t)$ is given  by
\begin {equation}
{\cal W}(t) = \left ({4\int_0^t 
%\left [
{\cal D}_F(t')
%+\epsilon E {\cal A}(t') \right ] 
\exp \left [ 2 \int^{t'} {\cal A}(t'')\, dt'' \right ]\, dt'}\right )
\left [{\exp \left (-2 \int_0^t {\cal A}(t')\, dt'\right )} \right ] \, \, .
\label{gauswid1}
\end{equation}

We further clarify that for a given initial condition, $\delta(p-p_0)$, 
and for the momentum independence approximation\footnote{It is
interesting to note that two of the present authors in Ref.~\cite{mmp} 
also used the same approximation and the said solution for a chemically 
equilibrating quark-gluon plasma in the same context~\cite{mp}.} 
of ${\cal A}$ and  ${\cal D}_F$
the above solution is correct and unique which could
easily be checked as the detailed calculational steps are given in 
Ref.~\cite{mt}. It is also worthwhile to note that for massless quarks
the analytically obtained solution of the FP equation in (\ref{fp}) 
agrees well with the solutions~\cite{gale} if one solves (\ref{fp0}) 
numerically for a given temperature ($T=400$ MeV) and initial momentum
($p_0=16$ GeV/c), up to $t\sim 8$ fm/c, beyond which the classical
approximations for transport coefficients are not valid. This
time regime is quite appropriate to study the energy loss probability 
distribution of a jet in quark-gluon plasma (QGP) whose life time 
is of the same order for RHIC energies.

Now, the authors of Ref.~\cite{mmp} claim that there should be also a linear 
term in momentum, $(p-\langle p\rangle)$, in the exponential of (\ref{solmom}) 
for reducing to the massless Maxwellian form  in the thermodynamic limit. 
Within the given initial condition, $\delta(p-p_0)$, and the momentum 
independence approximation of ${\cal A}$ and 
${\cal D}_F$, such a linear term in the exponent 
of (\ref{solmom}) does not appear (see also Lacuna 4).
Of course, if one changes the 
initial condition or relaxes the momentum independence of ${\cal A}$ and  
${\cal D}_F$, one will then arrive at an altogether different solution. 
Note that the authors of Ref.~\cite{mmp} have not defined
clearly what they really mean by $``$the possibility of 
a  different type of solution". However, we intend to discuss various
possibilities later in lacuna 4 while analysing the asymptotic forms
of the FP equation in detail.

\vspace{0.2in}

{\underline{\it Lacuna 3: Probability function in $p$ vs. $E$ }}

\vspace{0.2in}

According to (\ref{solmom}) the function $D(p,t)$ is a probability 
distribution in $p$ 
in one dimension ($-\infty \leq p \leq \infty$) for a given time $t$. 
Since it is an even function in $p$ the normalisation condition
can be written as 
\begin{equation}
\int_{-\infty}^{\infty}D(p,t) dp = 2\int_{0}^{\infty}D(p,t) dp = 
 2\int_{0}^{\infty}f(E,t) dE = 1 ,  \label{norm0} 
\end{equation}
with $E=p=|p|$. Thus, the normalisation requirement, for the massless case, is 
unambiguously preserved. This also holds for charm quarks as long
as we consider momenta $p$ much larger than the charm quark mass.

\vspace{0.2in}

{\underline{\it Lacuna 4: Unphysical asymptotic form}}

\vspace{0.2in}

Let us make another  $`$thought situation' where the drag (${\cal A}$) and 
the diffusion (${\cal D}_F$), are time independent, i.e., the test particle 
will experience a constant amount of drag and diffusion over a given length 
of medium or time traversed. This consideration is, however, without any 
loss of generality for the case of a static QGP.
 
In such a static medium the distribution in (\ref{solmom}) 
reduces to
\begin{equation}
D(p,t) = \sqrt{\frac{{\cal A}}{2\pi {\cal D}_F \left(1-e^{-2{\cal A}t}
\right)}} \ 
e^{-\frac{A{(p-p_0e^{-{\cal A}t})^2}}{2 {\cal D}_F
\left(1-e^{-2{\cal A}t}\right)}} \ \ . \label{fixedt}
\end{equation}
Now in the thermodynamic limit ($t\rightarrow \infty$), the above equation
simply becomes
\begin{equation}
D(p) = \sqrt{\frac{{\cal A}}{2\pi {\cal D}_F }} \ 
e^{-\frac{{\cal A}p^2}{2 {\cal D}_F}} \ \ . \label{thermo}
\end{equation}

At this point, we must check the uniqueness of the solution in (\ref{solmom}).
For the purpose, we refer to the second pair of the characteristic equation
in Ref.~\cite{mt}(See. as for example equation (22) in the first article 
in Ref.~\cite{mt})
corresponding to (\ref{fp}) in Fourier space, which amounts 
to the $t\rightarrow \infty$ limit, that reads 
\begin{equation}
\frac{\partial x}{{\cal A}x} \ = \ - \ \frac{\partial {\tilde D}}
{{\cal D}_Fx^2 {\tilde D}} ,  \label{second}
\end{equation}
where ${\tilde D}={\tilde D(x)}$ is the Fourier transform of the
momentum distribution function.
It is easy to show that the solution obtained using (\ref{second}) agrees 
perfectly with that of given in (\ref{thermo}). So, the solution obtained 
in Ref.~\cite{mt} 
is unique and a linear momentum term 
in the exponential does not arise, once the FP equation is given by
(C3) of Ref.~\cite{mmp} or by (\ref{fp}). 
However, we are aware of the fact that relaxing our 
momentum independence
approximation of ${\cal A}$ and ${\cal D}_F$, will certainly lead to  
different solutions, as the form of the FP equation will then be different 
from (\ref{fp}) or (C3) in Ref.~\cite{mmp}. We analyse various 
possibilities in the Appendix. 

Based on our analysis in the Appendix we  
note that the FP coefficients, $\Gamma_1$ and $\Gamma_2$, 
cannot simply be assumed in any form to ensure the correct form of 
the (equilibrium) distribution because this requires at least
an exact evaluation of the drag (${\cal A}$) and the diffusion 
(${\cal D}_F$) coefficients, which 
is even indeed a very involved task\footnote{Here, we do not even wish to 
speak of the higher order FP coefficients in the Taylor expansion.}~\cite{walton}. 
But knowing the collisional energy loss in terms of elementary collision 
reaction amplitudes, one can approximate the drag as in (\ref{drag}) and the 
diffusion as in (\ref{diffusion}).
The form of the approximation determines, a priori, the shape of the 
equilibrium distribution as discussed in the Appendix.  
Obviously, the momentum independence of ${\cal A}$ and ${\cal D}_F$ produces 
at least a consistent solution over the momentum range,
$-\infty\leq p \leq \infty$, in contrast to the other 
possibilities (see Appendix). 
Again,  we would also 
like to note that for small $t$ the change in diffusion\footnote{For 
example, see the second reference in Ref.~\cite{mt}.} 
will affect only
the width and the height of the distribution, but not the peak position which
is determined by the drag.
Since we really do not wish to obtain an equilibrium distribution for 
a jet but intend to obtain an energy-loss probability distribution over 
a finite length ($\sim$ 8 to 10 fm) of the medium, (\ref{solmom}) is a 
reasonably good approximation~\cite{mgm}.
In addition, it was independently 
verified~\cite{gale} numerically without any approximation as given 
in (\ref{fp0}), where  $\Gamma_1$ and $\Gamma_2$ are obtained using kinetic
theory calculations.  Nevertheless, it
would have been easier for us to discuss this problem if the authors of 
Ref.~\cite{mmp} would have been more specific in their statement
about $``$a  different type of solution". 

Now, the purpose of the study in Ref.~\cite{mt} was the following: before we 
took up these 
studies~\cite{mt}, the dominance of the radiative energy loss for the phenomenon
of jet quenching in heavy-ion collision seemed to be well established in the
heavy-ion community. Within
this simple approach we reconsidered, in 2003, the role 
of collisional partonic quenching and showed that it could be significant 
and  cannot just be overlooked as it were done in the literature. However, 
only recently, after publication 
of new data on the nuclear suppression factor in RHIC, this simple idea has 
gained wider interest. In fact, an additional contribution to the 
partonic energy loss appears to be necessary and a collisional component 
is a welcomed remedy, as advocated in~\cite{gyulassy}. At this point, since RHIC BNL
has provided very accurate data and LHC CERN will be operational soon,
one indeed needs to improve our simple approach in different possible ways, 
which would definitely 
be a very desirable to verify the importance of the collisional energy loss. 

\vspace{0.2in}

{\underline{\it Lacuna 5: Mean energy and loss }}

\vspace{0.2in}

According to our explanations on the above three points (Lacuna 2-4) 
the calculation of the mean energy obtained in Ref.~\cite{mt} 
follows from the expression of the distribution function itself in the classical
approximation. The mean momentum of the test particle, as for example we refer to
(\ref{fixedt}), is $\langle p \rangle=p_0e^{-{\cal A}t}$, along with the
diffusion process in momentum space as $\langle p^2 \rangle - 
\langle p \rangle^2 = 2\frac{{\cal D}_F}{{\cal A}}(1-e^{-2{\cal A}t})$. 

However, just to be sure, one can numerically compare  
$\langle p \rangle = p_0e^{-{\cal A} t}$
with the one 
given in (C11) of Ref.~\cite{mmp}. They agree up to three to five decimal 
places for time
interval $(1 - 10)$ fm/c as obviously the second term in (C11) is negligible
and our classical approximation is consistent. 
This suggests that using $\langle p \rangle =p_0e^{-{\cal A}t}$, 
is not a matter of great concern keeping 
in mind again the life-time of quark-gluon plasma. 
However, we note that it was used only for light quarks but proper 
care has been taken for heavy quarks.

\vspace{0.2in}

{\underline{\it Lacuna 6: Convolution integral for hadron spectrum}}

\vspace{0.2in}

We could not make out the point raised by the authors of Ref~\cite{mmp}
on our numerical calculations. In particular when they talk about a fixed 
length, $L$, in the convolution. However, the length $L(\phi)$ as 
defined in Ref.~\cite{mt} will be determined by a jet created at the 
transverse position, $r$ and the production angle, $\phi$ in central 
collision. 
With this $L$ one should convolute the spectra along with 
the geometry given by a cylinder of radius $R$ in Bjorken hydrodynamics,
which, however, should be restricted up to the critical temperature,
 $T_c\sim 0.2$ GeV.
We checked our numerical results and reproduced $Q(p_T)$ for light 
hadrons.  
The results obtained for the scaled energy loss, $\frac{\Delta E}{E}$, 
for heavy quarks have been independently verified in Ref.\cite{gyulassy}. 
Hence, we believe that the approach as well as the analytical 
and numerical solutions in Ref.\cite{mt} are correct as long as they are used 
within the domain of $(8-10)$ fm/c.

%\vspace{0.2in}

\begin{acknowledgments}
{ MGM gratefully acknowledges the financial support from
McGill India Strategic Research Initiative (MISRI) project during his
visit to Physics Department, McGill University. MGM is also thankful to 
Charles Gale, Sangyong Jeon, and Guang-You Qin for providing the numerical 
parts of the solution as well as for various fruitful discussions and suggestions,
and to Steffen Bass, Sanjay Ghosh, Rajarshi Ray and Dinesh K. Srivastava 
for critically 
reading the manuscript along with useful suggestions.} 
\end{acknowledgments}

\vspace{0.2in}

\appendix  

\section{Asymptotic Solution of Fokker-Planck Equation with various 
possibilities of its coefficients:}
%\label{other}

For convenience we write down the general form 
of the FP equation from (\ref{fp0}) for steady state as 
\begin{equation}
\frac{\partial}{\partial t}D(p,t) = \frac{\partial}{\partial p}
[\Gamma_1(p) D(p,t)] \ + \  \frac{\partial^2}{\partial^2 p}[{\Gamma_2}(p) 
D(p,t)] = 0 . \label{a0}
\end{equation}
where $D(p,t)$ is, in general, the probability distribution in momentum, 
$p \ (-\infty \leq p \leq \infty)$
at a given time $t \ (0\leq t \leq  \infty)$ of a test particle.
Now we investigate the asymptotic solutions of the FP equation with 
various  choices of its coefficients, $\Gamma_1(p)$ in (\ref{drag}) and 
$\Gamma_2(p)$ in (\ref{diffusion}).  

\vspace{0.2in}

{\underline{\it Type-I:}}

\vspace{0.12in}

We consider $\Gamma_1(p) \approx p{\cal A}$ and  
$\Gamma_2(p) \approx {\cal A}p^2$.
Now it is easy to show that the steady state 
solution of the FP 
equation in (\ref{a0}) reads as
\begin{equation}
D(p)= \frac{C_1}{p^3}   \ ,  \label{a1}
\end{equation}
where $C_1$ is a constant. This solution is not Maxwellian and diverges 
in the limit, $p \rightarrow 0$. Thus, it is not a consistent 
solution over $-\infty \leq p \leq \infty$.

\vspace{0.20in}

{\underline{\it Type-II:}}

\vspace{0.12in}

We now consider $\Gamma_1(p) \approx p{\cal A}$ and  
$\Gamma_2(p) \approx {\cal A}Tp$.
Similarly, the form of the steady state solution of the FP equation 
in (\ref{a0}) can be obtained as
\begin{equation}
D(p)= \frac{C_2}{p} \ e^{-\frac{p}{T}}   \ .  \label{a2}
\end{equation}
This appears like a pseudo-Maxwellian form but is again not a consistent 
solution over $-\infty \leq p \leq \infty$ as it diverges in the 
limit, $p\rightarrow 0$, like the Type-I above. 
However, it is very easy to anticipate the reason.

\vspace{0.2in}

{\underline{\it Type-III (our case):}}

\vspace{0.12in}

We consider $\Gamma_1(p) \approx p{\cal A}$ and  $\Gamma_2 = {\cal D}_F$.
The steady state solution is
\begin{equation}
D(p)= C_3 \ e^{-\frac{Ap^2}{2D_F}}   \ ,  \label{a3}
\end{equation}
which is, however, not of the desired Maxwellian form in usual sense 
but fully consistent over $-\infty \leq p \leq \infty$. 

\vspace{0.4in}
%\newpage

{\underline{\it Type-IV :}}

\vspace{0.2in}

We now explore the simplest one where both $\Gamma_1$ and $\Gamma_2$ 
are momentum independent. 
Then the steady state solution reads
\begin{equation}
D(p)= C_4 \ e^{-\frac{\Gamma_1}{\Gamma_2}p}   \ .  \label{a4}
\end{equation}
Now one can claim it to be of the Maxwellian form
just by choosing 
$\frac{\Gamma_1}{\Gamma_2}=\frac{1}{T}$.

\end{document}